\documentclass[11pt,a4paper]{article}

\usepackage[truedimen,margin=34mm]{geometry} 

\usepackage{mathrsfs}
\usepackage{amssymb}
\usepackage{amsmath}
\usepackage{ascmac}
\usepackage{amsthm}
\usepackage[pdftex]{graphicx}
\usepackage[pdftex]{color}
\usepackage{booktabs}
\usepackage{setspace}
\usepackage{natbib}
\usepackage{color}

\usepackage{comment}

\usepackage{titlesec}
\titleformat*{\section}{\large\bfseries}
\titleformat*{\subsection}{\it}

\newtheorem{thm}{Theorem}

\def\ep{{\varepsilon}}

\def\si{{\sigma}}
\def\th{{\theta}}

\def\sih{{\widehat \si}}

\def\muh{{\widehat \mu}}

\def\At{\widetilde{A}}
\def\tht{\widetilde{\th}}

\title{{\bf Parametric Bootstrap Confidence Intervals for the Multivariate Fay-Herriot Model}}

\date{}
\author{}

\begin{document}

\maketitle
\doublespacing

\vspace{-1.5cm}
\begin{center}
{\large Takumi Saegusa$^1$, Shonosuke Sugasawa$^3$ and Partha Lahiri$^{12}$}
\end{center}

\noindent
$^1$Department of Mathematics, The University of Maryland, College Park\\
$^2$Joint Program in Survey Methodology, The University of Maryland, College Park\\
$^3$Center for Spatial Information Science, The University of Tokyo

\medskip
\noindent

\vspace{5mm}
\begin{center}
{\bf \large Abstract}
\end{center}
The multivariate Fay-Herriot model is quite effective in combining information through correlations among small area survey estimates of related variables or historical survey estimates of the same variable or both. Though the literature on small area estimation is already very rich, construction of second-order efficient confidence intervals from multivariate models have so far received very little attention. In this paper, we develop a parametric bootstrap method for constructing a second-order efficient confidence interval for a general linear combination of small area means using the multivariate Fay-Herriot normal model. The proposed parametric bootstrap method replaces difficult and tedious analytical derivations by the power of efficient algorithm and high speed computer. Moreover, the proposed method is more versatile than the analytical method because the parametric bootstrap method can be easily applied to any method of model parameter estimation and any specific structure of the variance-covariance matrix of the multivariate Fay-Herriot model avoiding all the cumbersome and time-consuming calculations required in the analytical method. We apply our proposed methodology in constructing confidence intervals for the median income of four-person families for the fifty states and the District of Columbia in the United States. Our data analysis demonstrates that the proposed parametric bootstrap method generally provides much shorter confidence intervals compared to the corresponding traditional direct method. Moreover, the confidence intervals obtained from the multivariate model is generally shorter than the corresponding univariate model indicating the potential advantage of exploiting correlations of median income of four-person families with median incomes of three and five person families.

\bigskip\noindent
{\bf Key words}: Empirical Best predictor; higher-order asymptotics; small area estimation.

\newpage
\section{Introduction}
For the last few decades, there has been an increasing demand to produce reliable estimates for small geographic areas, commonly referred to as small areas, since such estimates are routinely used for fund allocation and regional planning. The primary data, usually a survey data, are usually too sparse to produce reliable direct small area estimates that use data from the small area under consideration. To improve upon  direct estimates, different small area estimation techniques that use multi-level models to combine information from relevant auxiliary data have been proposed in the literature.  The readers are referred to  \cite{JL2006} and \cite{Rao2015} for a comprehensive review of small area estimation.

In estimating per-capita income of small places (population less than 1000), \cite{FH1979} proposed an empirical Bayes method to improve on direct survey-weighted estimates by borrowing strength from administrative data and survey estimates from a bigger area.  Their method uses a two-level normal model in which the first level captures the variability of the survey estimates and the second level links the true small area means to aggregate statistics from administrative records and survey estimates for a bigger area.  Researchers working on small area estimation have found the Fay-Herriot model useful in investigating various theoretical properties as well as implementing methodology in different applied problems when we do not have access to micro-data because of confidentiality and other reasons.  For a review on the Fay-Herriot model and the related empirical best predictions, readers are referred to \cite{Lahiri2003a}.

Following the pioneering paper by \cite{FH1979}, several multivariate extensions of the Fay-Herriot model have been considered to  combine information from small area estimates of related variables or from past small area estimates of the same variable or both. They are essentially special cases of the general multivariate random effects or two-level multivariate model.  In the context of estimating median income of four-person families for the fifty states and the District of Columbia (small areas), \cite{Fay1987} suggested a multivariate extension of the Fay-Herriot model, commonly referred to as the multivariate Fay-Herriot model, in order to borrow strength from the corresponding survey estimates of median income of three and five person families for the small areas. Alternatively, in Fay's setting one could think of using survey estimates of median income for the three-person and five-person families as auxiliary variables in an univariate Fay-Herriot model.  But, unlike the univariate Fay-Herriot model, the multivariate Fay-Herriot model incorporates sampling variance-covariance matrix of direct survey estimates of median income of the 3-, 4- and 5-person families for each small area.  Moreover,  the multivariate Fay-Herriot model borrows strengths through correlations of the components of area specific vector of random effects associated with the true median income of 3-, 4-, and 5- person families.  Inferences on the four-person median income for the small areas drawn from the multivariate Fay-Herriot model are expected to be more efficient and reasonable when compared to the inferences drawn from an univariate Fay-Herriot model with survey estimates of the median income for three and five person families as auxiliary variables.  This is because the univariate Fay-Herriot model would ignore the sampling variability of the survey estimates of median income for the three and five person families in the small areas.

In estimating median income of four-person families for the fifty states and the District of Columbia, \cite{DFG1991} used a bivariate Fay-Herriot model with a general structure for the variance-covariance matrix of the  vector of area specific random effects. However, in many small area applications, structured variance-covariance matrices for the vector of area specific random effects arise naturally.  For example,  in order to combine information from the related past data,   \cite{RY1994} proposed a stationary time series cross-sectional model while \cite{DLM2002} proposed a random walk time series and cross-sectional model.   Although the time series cross-sectional models can be viewed as special cases of the multivariate Fay-Herriot model, one can achieve greater efficiency in estimating the unknown variance covariance matrix by reducing the number of parameters in the variance-covariance matrix through time series cross-sectional models.

Empirical best linear unbiased predictions and associated uncertainty measures for multivariate Fay-Herriot models with or without structured variance-covariance matrices for the vector of random effects have been adequately studied; see, e.g., \cite{DFG1991}, \cite{RY1994}, \cite{BM2016}, \cite{DLM2002}, and others.  However, the problem of constructing second-order efficient confidence intervals for the multivariate Fay-Herriot model, i.e., confidence intervals with coverage error $o(m^{-1}),$ $m$ being the number of small areas, received very little attention. 
 \cite{Datta2002} obtained  a second-order efficient confidence interval for a small area mean using an analytical method. To this end, they first obtained the exact expression for the term of order $O(m^{-1})$ in a higher order expansion of coverage probability of a normality-based empirical Bayes confidence interval, originally proposed by \cite{Cox1975}, and then, using the $O(m^{-1})$ term in the expansion, suggested an adjustment to the normal percentile in order to lower the coverage error to $o(m^{-1})$. The approach of \cite{IK2018} in obtaining a second-order efficient confidence region for the vector of means for each area is essentially a multivariate generalization of \cite{Datta2002}.  However, their results are specifically designed for  the multivariate Fay-Herriot model with an unstructured variance-covariance matrix when a method-of-moment estimator of the variance-covariance matrix of the vector of random effect is used.  The derivation of the second-order efficient confidence intervals by the analytical method of \cite{Datta2002} or \cite{IK2018} is cumbersome and one needs to go through the such derivation each time one changes the model (say, a multivariate Fay-Herriot model with model variance-covariance structure suggested by the time series cross-sectional model of \cite{RY1994} or \cite{DLM2002}) or estimation method for the model parameters.


Parametric bootstrap method for obtaining second-order unbiased mean squared error estimation was first proposed by \cite{BL2002}. Construction of the second-order efficient confidence interval based on the empirical best linear predictor of a small area parameter for a general linear mixed model was proposed by \cite{CLL2008}.  For parametric bootstrap confidence intervals for the univariate Fay-Herriot model, see \cite{Lahiri2003} and \cite{LL2010}. In this paper, we develop a parametric bootstrap method for obtaining  second-order efficient confidence intervals for small area parameters from a multivariate Fay-Herriot model. Compared to the analytical method, our parametric bootstrap approach for constructing second-order confidence intervals for small area parameters  is versatile and theoretically complete because our method applies to any variance estimator with minimal assumptions and theoretical justification is directly provided to the proposed method.

In section 2, we describe the multivariate model, associated estimation of the model parameters, and the proposed parametric confidence interval for a linear combination of small area means. We present our data analysis in section 3.  An outline of the technical proof of our main result is deferred to the Appendix.

\section{Parametric Bootstrap Confidence Intervals for the Multivariate Fay-Herriot Model}

\subsection{Multivariate Fay-Herriot model}
Let $\th_i=(\th_{i1},\ldots,\th_{is})$ and  $y_i=(y_{i1},\ldots,y_{is})$ be a vector of characteristics of interest and a vector of direct survey estimates of $\th_i$ for area $i,\;(i=1,\ldots,m)$, respectively, where $m$ is the number of small areas. 
The multivariate Fay-Herriot model \citep{Fay1987, BM2016} is given by 
\begin{equation}\label{MFH}
y_i=\th_i+\ep_i, \ \ \ \th_i=X_i\beta+v_i, \ \ \ i=1,\ldots,m
\end{equation}
where $X_i$ is a $s\times p$ matrix of known explanatory variables; $\ep_i$ and $v_i$ are vectors of area specific sampling errors and random effects, respectively; $\{\ep_i,\;i=1,\ldots,m\}$ and $\{v_i,\;i=1,\ldots,m\}$ are all independent with $\ep_i\sim N(0,D_i)$ 
and $v_i\sim N(0,A(\psi))$, $D_i$ being the $s\times s$ known sampling variance-covariance matrix of $y_i,\;i=1,\ldots,m$.

We assume that  $A(\psi)$, the variance-covariance matrix of the random effects $\theta_i$,  depends on $k$ unknown parameters  $\psi=(\psi_1,\ldots\psi_k)$ with $1\leq k\leq s(s+1)/2$.  For the small area application considered by \cite{DFG1991}, $A(\psi)$ is an unstructured variance-covariance matrix with $s=2$ and $k=3$.  For the stationary time series cross-sectional model of \cite{RY1994}, $A(\psi)$ is a structured variance-covariance matrix with $s$ as the number of time points, and $k=3.$  For the random walk time series cross-sectional model of \cite{DLM2002}, $A(\psi)$ is a structured variance-covariance matrix with $s$ as the number of time points, and $k=2.$  Let $\phi=(\beta,\psi)$ be a vector of all the unknown parameters.

For unified representations over $m$ areas, we define $y=(y_1^t,\ldots,y_m^t)^t$, $X=(X_1^t,\ldots,X_m^t)^t$, and define $v, \ep$ and $\th$ in the same way as $y$. 
Then, the model can be expressed as
\begin{equation*}y=X\beta+v+\ep,
\end{equation*}
where $v\sim N(0,\At(\psi))$ with $\At(\psi)={\rm diag}(A(\psi),\ldots,A(\psi))\in \mathbb{R}^{ms\times ms}$ and $\ep\sim N(0,D)$ with $D={\rm diag}(D_1,\ldots,D_m)\in \mathbb{R}^{ms\times ms}$.
With this notation, we can write 
${\rm Var}(y)\equiv \Sigma={\rm diag}( A(\psi)+D_1,\ldots,A(\psi)+D_m)$.
In this paper, we are interested in constructing  confidence intervals for $T=c^t\theta$, where $c$ is a $ms$-dimensional vector of known constants. 
For example, if we let $c=(1,0,\ldots,0)$, $T=\th_{i1}$ is the first characteristics in the first area, and $T$ also can be the difference of characteristics in different areas by setting $c$ appropriately.

Under the model (\ref{MFH}), the best linear unbiased predictor of $\theta_i$ with known parameters is given by 
$$
\tht_i=y_i-D_i\{A(\psi)+D_i\}^{-1}(y_i-X_i\beta), \ \ \ i=1,\ldots,m, 
$$
which shrinks $y_i$ toward the regression part $X_i\beta$.
Note that each element in $\tht_i$ depends not only on the corresponding observation but also other observations in the same area when $D_i$ or $A(\psi)$ have non-zero off-diagonal elements. 
Exploiting the information on the  correlation structure, the best linear unbiased predictor would be able to provide more accurate estimates of $\th_i$ than simple applications of the univariate FH models to each element.
In fact, it will be numerically shown that such advantage is inherited to interval lengths of confidence intervals. 
The multivariate Fay-Herriot model provides more efficient confidence intervals than the univariate Fay-Herriot model by borrowing information from related components.

\subsection{Estimation of model parameters }
Because the best linear unbiased predictor $\tht_i$ depends on unknown parameters, statistical inference on $\theta_i$ is carried out via the empirical best linear unbiased estimator given by
\begin{equation*}
    \widehat{\theta}_i = y_i-D_i\{A(\widehat{\psi})+D_i\}^{-1}(y_i-X_i\widehat{\beta})
\end{equation*}
where $A(\widehat{\psi})$ and $\widehat{\beta}$ are some estimators of $A(\psi)$ and $\beta$.
We estimate $\beta$ by the generalized least squares estimator 
\begin{equation*}
    \widehat{\beta} = (X^t\widehat{\Sigma}^{-1}X)^{-1}X^t\widehat{\Sigma}y
\end{equation*}
once $A(\widehat{\psi})$ in $\widehat{\Sigma}$ is obtained.
There are several different methodology to estimate $A(\psi)$ (e.g.\@ the restricted maximum likelihood estimator \citep{BM2016} and moment-based estimators \citep{IK2018}), but the proposed method to construct the empirical Bayes confidence interval does not depend on a  specific variance estimator.
For the data analysis below, 
we adopt the maximum likelihood estimator that maximizes
$$
L(\phi)=-\frac12\sum_{i=1}^m\log |A(\psi)+D_i|-\frac12\sum_{i=1}^m(y_i-X_i\beta)^t\{A(\psi)+D_i\}^{-1}(y_i-X_i\beta)
$$
by the EM algorithm.
Note that this method estimates $\beta$ and $A(\psi)$ simultaneously and automatically yields the generalized least squares estimator $\widehat{\beta}$ as the maximum likelihood estimator.

\subsection{Confidence intervals via parametric bootstrap}
We describe our methodology to construct the empirical Bayes confidence interval for $T=c^t\theta$. 
To motivate our method, we first consider a traditional approach to interval estimation.
The key observation for this approach is that 
the conditional distribution of $T=c^t\theta$ under the model (\ref{MFH}) is given by $T|y\sim N(\mu_T, \si_T^2)$, where 
\begin{eqnarray*}
&&\mu_T\equiv \mu_T(y,\phi)=c^tD\Sigma^{-1}X\beta+c^t\At(\psi)\Sigma^{-1}y, \\
&&\si_T^2\equiv\si^2_T(\psi)=c^t{\rm diag}\left((A(\psi)^{-1}+D_1^{-1})^{-1},\ldots,(A(\psi)^{-1}+D_m^{-1})^{-1}\right)
c.
\end{eqnarray*}
Since $\si_T^{-1}(T-\mu_T)$ follows the standard normal distribution, one can find $z$ such that $P(\si_T^{-1}|T-\mu_T|\leq z)=1-\alpha$ for a fixed $\alpha\in (0,1)$. Because the resultant interval $(\mu_T\pm z\sigma_T)$ for $T$ contains unknown parameters $\mu_T$ and $\sigma_T$, the traditional approach replaces these parameters by their consistent estimators $\widehat{\mu}_T$ and $\widehat{\sigma}_T$ to obtain the confidence interval $(\widehat{\mu}_T\pm z\widehat{\sigma}_T)$ for $T$. 
Though this interval has a correct coverage asymptotically, it tends to be too short or too long in practice.
This undesirable phenomenon is due to the reliance on the rather crude approximation of the standard normal distribution by $\sih_T^{-1}(T-\muh_T)$, which yields the coverage error of $O(m^{-1})$.
Because $\mu_T$ and $\sigma_T$ must be estimated, the issue of the asymptotic approximation is not avoidable. Instead, we consider the distribution of  $\sih_T^{-1}(T-\muh_T)$ from the beginning and consider a method to precisely approximate it.
We achieve this goal through the parametric bootstrap.

We construct the bootstrap sample in a prametric way as follows. 
First we independently generate $v_i^*\sim N(0,A(\widehat{\psi}))$ and $\epsilon_i^*\sim N(0,D_i)$. 
Because $\theta_i = X_i\beta+v_i$ and $y_i=\theta_i+\epsilon_i$ in the model (\ref{MFH}),
we construct 
\begin{eqnarray*}
\theta^*_i &=& X_i\widehat{\beta}+v_i^*,\\
    y_i^* &=& \theta_i^*+\epsilon_i^*.
\end{eqnarray*}
The resultant bootstrap sample is  $\{(y_1^*,X_1),\ldots,(y_m^*,X_m)\}$.
To approximate $\sih_T^{-1}(T-\muh_T)$, we compute $T^*=c^t\theta^*$ with $\theta^*=((\theta_1^*)^t,\ldots,(\theta_m^*)^t)^t$.
Bootstrap estimators $\widehat{\mu}_T^*$ and $\widehat{\sigma}^*$ of $\mu_T$ and $\sigma_T$ is obtained in the same way as $\widehat{\mu}_T$ and $\widehat{\sigma}_T$ by replacing the original sample $y_i$ by the bootstrap sample $y_i^*$.
For example, one can compute the bootstrap maximum likelihood estimator $\widehat{\phi}^* = (\widehat{\beta}^*,\widehat{\psi}^*)$ by maximizing
 $$
L^*(\phi)=-\frac12\sum_{i=1}^m\log |A(\psi)+D_i|-\frac12\sum_{i=1}^m(y_i^*-X_i\beta)^t\{A(\psi)+D_i\}^{-1}(y_i^*-X_i\beta)
$$
as in our data analysis below. Once $\widehat{\phi}^*$ is computed, we plug this in to obtain 
$\widehat{\mu}_T^* = \mu_T(y^*,\widehat{\phi}^*)$
and $\widehat{\sigma}_T^* = \sigma_T(\widehat{\psi}^*)$.

The conditional distribution of 
$$\sih_T^{-1\ast}(T^{\ast}-\muh_T^{\ast})$$
given the data $y$ is the parametric bootstrap approximation of the distribution of $\sih_T^{-1}(T-\muh_T)$.
Because the random variable $\sih_T^{-1\ast}(T^{\ast}-\muh_T^{\ast})$ can be generated as described above, one can find the quantity $(q_1,q_2)$ that satisfies $P(q_1\leq \sih_T^{-1\ast}(T^{\ast}-\muh_T^{\ast})\leq q_2) = 1-\alpha$ as precisely as possible. Because parametric bootstrap provides a precise approximation, $(q_1,q_2)$ is expected to yield a similar probability for  $\sih_T^{-1}(T-\muh_T)$.
The proposed parametric bootstrap confidence interval is 
\begin{equation*}
 \widehat{\mu}_T +\widehat{\sigma}_Tq_1\leq    T\leq \widehat{\mu}_T +\widehat{\sigma}_Tq_2.
\end{equation*}
The following theorem states that the proposed empirical Bayes confidence interval achieves correct coverage asymptotically with error $O(m^{-3/2})$.

\begin{thm}\label{thm:main}
We assume the following conditions: 
\begin{itemize}
\item 
The matrix $X$ is of full rank satisfying
$(X^t\Sigma^{-1} X)^{-1} =O(m^{-1})$.

\item
$A(\widehat{\psi})$ is a strictly positive definite matrix satisfying 
$ E\lVert A(\widehat{\psi})-A(\psi)\rVert_F=O(m^{-1})$ where
$\lVert \cdot \rVert_F$ be the Frobenius norm.

\item
There exists positive constants $\underline{\lambda}$ and $\overline{\lambda}$ such that the sampling variance-covariance matrix $D_i$ satisfies $\underline{\lambda}I_s \leq D_i \leq \overline{\lambda}I_s$ for $i=1,\ldots,m$.
\end{itemize} 
Let $\alpha\in (0,1)$.
Suppose $(q_1,q_2)$ satisfies
\begin{equation*}
    P(q_1\leq \sih_T^{-1\ast}(T^{\ast}-\muh_T^{\ast})\leq q_2) = 1-\alpha.
\end{equation*}
Then 
$$
P\left(\widehat{\mu}_T +\widehat{\sigma}_Tq_1\leq    T\leq \widehat{\mu}_T +\widehat{\sigma}_Tq_2\right)=1-\alpha+
O(m^{-3/2}).
$$
\end{thm}

\section{Application}\label{sec:app}
In this section, we use old data used earlier by \cite{DFG1991} to compare three different confidence interval methods: direct method, parametric bootstrap confidence interval methods -- one based on an univariate Fay-Herriot model and the other based on multivariate Fay-Herriot model. The data contain direct survey estimates  of median income of 3-, 4- and 5-person families and their associated standard errors for the fifty states and the District of Columbia during years 1979-88. In addition, data contain census median income of 3-, 4-, and 5-person families obtained from the 1970 and 1980 decennial censuses. The U.S. Department of Health and Human Services (HHS) administers a program of energy assistance to low-income families. Eligibility for the program is determined by a formula where the most important variable is an estimate of the current median income for four-person families by states.

Let $\theta_{i1}, \theta_{i2}$ and $\theta_{i3}$ denote the true median income of 3-, 4- and 5-person families, respectively, for $i=1,\ldots,m$, where $m=51$ is the number of states and the District of Columbia in the United States.
Let $y_{i1}, y_{i2}$ and $y_{i3}$ be the corresponding direct survey estimates.
Our primary interest is the four-persons family median income, $\theta_{i2}$, and we consider estimating the parameter by borrowing strength from not only area specific auxiliary variables but also from the direct survey estimates of median income for the 3- and 5-person families. As for the area specific auxiliary variables, we consider the median income data obtained from the most recent decennial census and an 'adjusted' census median income obtained by multiplying the most recent census median income by the ratio of per-capita income of the current year to the most recent decennial census year. The per-capita income information is available from administrative records maintained by the Bureau of Economic Analysis (BEA). 
Then, the covariate matrix $X_i$ is a $3\times 9$ matrix given by
$$
X_i={\rm diag}((1, x_{i1}, x_{i1}^{\ast}), (1, x_{i2}, x_{i2}^{\ast}), (1, x_{i3}, x_{i3}^{\ast})),
$$
where $x_{ik}$ and $x_{ik}^{\ast}$ denote the census data and adjusted census in the $i$th area for three-person $(k=1)$, four-family $(k=2)$ and five-person $(k=3)$ family median incomes.

We first applied the multivariate Fay-Herriot model (MFH) given in (\ref{MFH}) separately to the survey data in each year from 1981 to 1988, where we used median income of 1979 obtained from the 1980  decennial census data as auxiliary variables.
For comparison, we also applied the univariate Fay-Herriot (UFH) model only to the four-person family income data $y_{i2}$ with the corresponding census data as auxiliary variables.
We found that the maximum likelihood estimates of the random effects variance in the UFH model were 0 in 1982, 1983 and 1986, in which confidence intervals of $\theta_{i2}$ cannot be obtained. 
On the other hand, we observed that the MFH model produces positive definite estimates for $A$ in all the years, and correlations are quite high in some years.
This indicates that the random effects variance in $\theta_{i2}$ can be stably estimated by borrowing strength from other information such as $y_{i1}$ and $y_{i3}$ through the MFH model (\ref{MFH}).
For illustration, we focus on the results in 1984 and 1987 in which the estimated correlation matrices are given by 
$$
\left(\begin{array}{ccc}
  1   & 0.171 & 0.938 \\
  0.171 & 1  & 0.200 \\
  0.938 & 0.200 & 1
\end{array}\right), \ \ \ \ \ \ 
\left(\begin{array}{ccc}
  1   & 0.780 & 0.587 \\
  0.780 & 1  & 0.915 \\
  0.587 & 0.915 & 1
\end{array}\right),
$$
respectively. 
Note that the correlations are quite high in 1987 while relatively small in 1984.

\begin{figure}[htb!]
\centering
\includegraphics[width=14cm,clip]{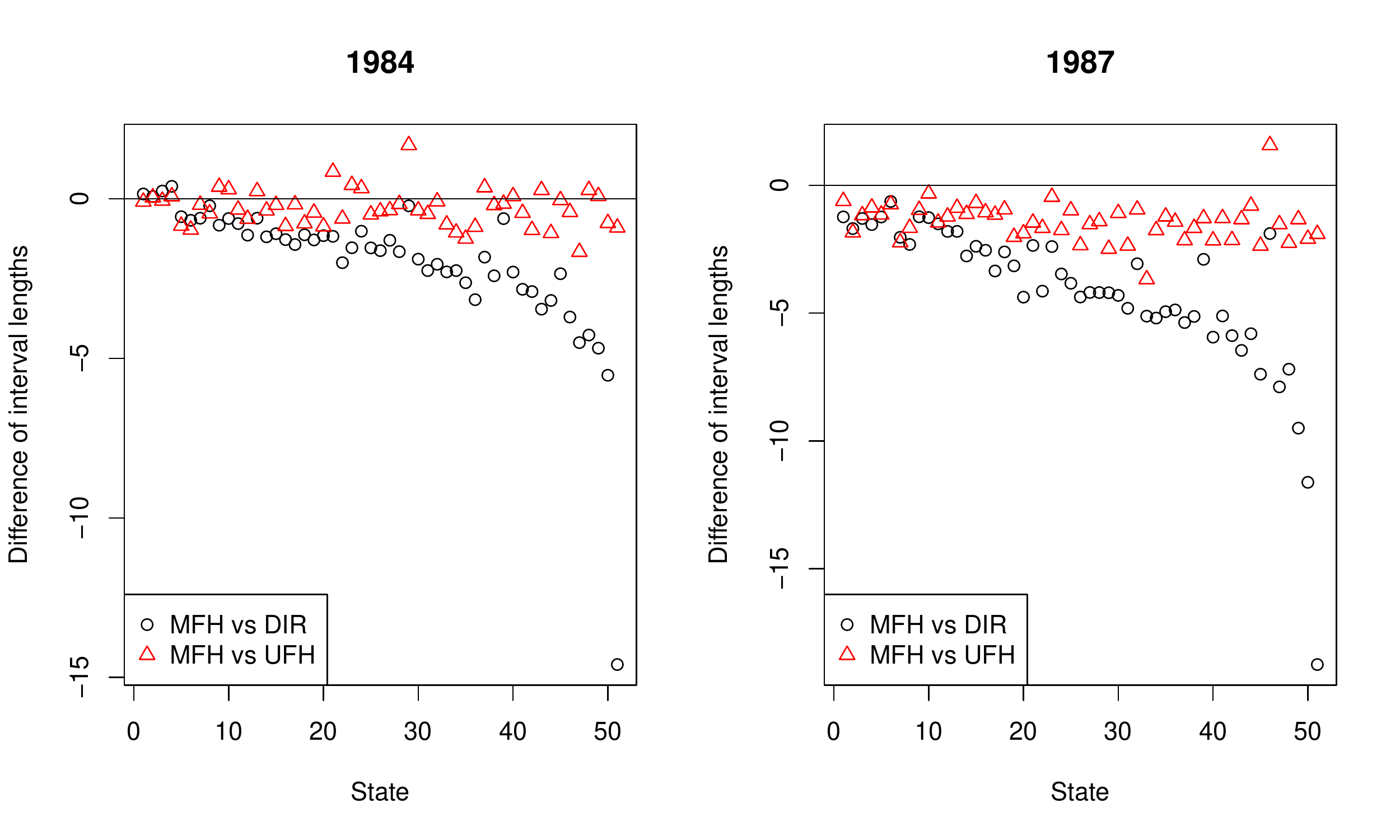}
\caption{Differences of 95$\%$ confidence interval lengths based on the multivariate Fay-Herriot (MFH), univariate Fay-Herriot (UFH) and the naive method with direct estimator (DIR) in 1984 and 1987 surveys.
The states are arranged in the ascending order with respect to the sampling variances.}
\label{fig:CI}
\end{figure}

Based on 1000 bootstrap replications, we computed 95$\%$ confidence intervals of $\theta_{i2}$ under both MFH and UFH models.
We also computed 95$\%$ confidence intervals based on the direct estimator (denoted by DIR), given by $(y_{i2}-z_{0.025}\sqrt{D_{i22}}, y_{i2}+z_{0.025}\sqrt{D_{i22}})$, where $z_{0.025}$ is the upper $0.025$ quantile of the standard normal distribution, and $D_{i22}$ is the $(2,2)$-element of $D_i$.
In Figure \ref{fig:CI}, we present the differences in lengths of 95$\%$ confidence interval based on the MFH model, the UFH model and the DIR method, where the states are arranged in the ascending order of sampling variances. 
Negative values of the difference indicate that the lengths of confidence intervals from the MFH model are shorter than those from the UFH model or the DIR method.
We also reported summary values of area-wise confidence intervals in Table \ref{tab:CI-summary}.
Comparing the inverval from the MFH model and the DIR method, the difference tends to be larger as the sampling variance increases. This is reasonable because we can improve the accuracy of inference on parameters in areas with large sampling variance by borrowing strength through the model. 
Comparing intervals from the MFH and UFH models, two lengths are comparable in 1984 possibly because the correlations among three median incomes are not so strong. The advantage of borrowing strength from the other incomes can be limited.
On the other hand, in 1987, the MFH model produces shorter confidence intervals than the UFH model in almost all the areas due to the high correlations.

We next investigated the performance of the confidence intervals using the census data in 1979 as if they were true values.
We applied the MFH and UFH methods to the survey data in 1979 using 1969 census data as covariates.
In this case, we applied the UFH method to not only $y_{i2}$ but also $y_{i1}$ and $y_{i3}$.
We first found that in the UFH model  both maximum likelihood and restricted maximum likelihood estimates of the random effects variances are zero (for $y_{i2}$ and $y_{i3}$) or very small (for $y_{i1}$).
Based on 1000 bootstrap replications and the maximum likelihood method, we obtained $95\%$ confidence intervals of $\theta_{ik}$ with $k=1,2,3$, in the MFH and UFH models.
We calculated mean and median lengths of area-wise confidence intervals, denoted by Len1 and Len2, respectively. 
We also computed the empirical coverage rate (CR) by considering 1979 census data as true values.
The results are reported in Table \ref{tab:validation}.
Although the UFH model provides shorter confidence intervals than the MFH model, the empirical coverage rate is quite low compared with the nominal level $95\%$.
This suggests that the confidence intervals in the UFH model are too liberal in this case, possibly because of the small estimates of random effects variance. 
On the other hand, the MFH model provides reasonable confidence intervals.
Their coverage rates are quite high and their lengths are much shorter than the those in the direct method.

\begin{table}[!htb]
\caption{Summary measures of interval lengths of three methods. 
\label{tab:CI-summary}
}
\begin{center}
\begin{tabular}{cccccccccc}
\hline
Year & Method  & min & 25$\%$ & Median & Mean & 75$\%$ & max \\
 \hline
 & MFH & 4.12 & 5.55 & 6.13 & 6.02 & 6.38 & 8.28 \\
1984 & UFH & 4.21 & 5.78 & 6.42 & 6.32 & 6.72 & 8.44 \\
 & DIR & 3.96 & 6.66 & 7.75 & 8.00 & 8.84 & 21.30 \\
 \hline
 & MFH & 4.54 & 6.08 & 6.78 & 6.74 & 7.23 & 12.87 \\
1988 & UFH & 5.86 & 7.60 & 8.57 & 8.39 & 8.95 & 11.85 \\
 & DIR& 6.18 & 8.23 & 10.71 & 10.84 & 12.15 & 32.39 \\
 \hline
\end{tabular}
\end{center}
\end{table}

\begin{table}[!htb]
\caption{Performance of $95\%$ confidence intervals of MFH, UFH and DIR.  
The results of UFH in four- and five-persons family incomes do not exist because of zero estimates of the random effects variance.
\label{tab:validation}
}
\begin{center}
\begin{tabular}{cccccccccccccccccc}
\hline
 && \multicolumn{3}{c}{three-persons family} &&  \multicolumn{3}{c}{four-persons family} && \multicolumn{3}{c}{five-persons income}\\
Method  & & CR & Len1 & Len2  & & CR & Len1 & Len2 & & CR & Len1 & Len2  \\
 \hline
MFH &  & 100 & 3.21 & 3.13 &  & 96.1 & 2.40 & 2.29 &  & 100 & 4.52 & 4.36 \\
UFH &  & 74.5 & 1.77 & 1.69 &  &  -  &  -  &  -  &  &  -  &  -  &  -  \\
DIR &  & 86.3 & 5.57 & 5.41 &  & 86.3 & 5.88 & 5.91 &  & 86.3 & 9.25 & 8.98 \\
 \hline
\end{tabular}
\end{center}
\end{table}

\section{Concluding Remarks}
In this paper, we proposed the parametric bootstrap method for computing a second-order accurate confidence interval of 
small area parameters from the multivariate Fay-Herriot model.
The proposed parametric bootstrap method is easy to implement and is widely applicable to many variance estimators with minimal assumptions as seen in Theorem \ref{thm:main}.
This advantage forms a sharp contrast to the analytical calibration proposed by \cite{Datta2002} and \cite{IK2018} where a different estimation method of model parameters requires cumbersome derivations of the correction terms and tedious checking of assumptions.
We demonstrated the superior performance of the proposed methodology over the univariate and direct methods in the family income data.
Better coverage and generally shorter length of the proposed interval is due to the effective use of the correlation structure in the same area and direct approximation of the distribution through parametric bootstrap.

There are several future directions to extend the proposed methodology.
An immediate extension is to construct the confidence region of a vector of small area parameters  studied by \cite{IK2018}. In the current paper, we focused on the linear combination of small area parameters because the primary interest lies in the single parameter or difference in two parameters in practice. 
When more than three parameters are of interest, our simple and versatile parametric bootstrap method is expected to be a powerful alternative to the analytical calibration. Though  methodology itself is exactly the same as in the current paper,  theoretical justification of parametric bootstrap is a challenging problem because it involves multivariate integrals.
Another direction is to extend the parametric bootstrap to a more general multivariate linear mixed models.
Because theoretical arguments by \cite{CLL2008} for the general univariate case is similar to that in this paper, its multivariate extension can be done in a similar way.
Another interesting question is to address the issue of non-positive definiteness of the estimated variance-covariance matrices.
Singularities of estimated variance-covariance matrices may occur both in the original estimate and the bootstrap estimate.
Because this issue compromises the validity of the parametric bootstrap procedure, 
it is important to develop reasonable adjustment methods possibly motivated from existing approaches in the univariate situation \citep[e.g.][]{LL2010}.

\section*{Acknowledgement}
The second author's research was partially supported by Japan Society for Promotion of Science (KAKENHI) 18K12757.
The third author's research was partially supported by the U.S. National Science Foundation Grants SES-1758808.

\vspace{1cm}
\appendix
\begin{center}
{\bf Appendix}
\end{center}

For notational simplicity, we suppress the dependence on $\psi$. For example, we write $A$ and $\widehat{A}$ for $A(\psi)$ and $A(\widehat{\psi})$. 
\section*{Proof of Theorem}

Recall that the conditional distribution of $\theta$ given $Y$ is the multivariate
normal distribution with mean $\mu=(\mu_1^t,\ldots,\mu_m^t)^t$ and the 
variance-covariance matrix $\sigma^2={\rm diag}\{(A^{-1}+D_1^{-1})^{-1},\ldots,(A^{-1}+D_m^{-1})^{-1}\}$ where $\mu_i = A(A+D_i)^{-1}y_i+ D_i(A+D_i)^{-1}X_i\beta$.
Let $T = c^t\theta$. The conditional distribution of $T$ given $Y$ is then the
normal distribution with mean $\mu_T = c^t\mu$ and variance
$\sigma_T^2 =c^t\sigma^2c$.
Let $\Phi$ and $\phi$ be the cumulative distribution function and density function for the standard normal random variable.
Define $Q(Y)=\sigma_T^{-1}\left\{\muh_T-\mu_T+r(\sih_T-\si_T)\right\}$.
It follows that 
\begin{align*}
P(\sih_T^{-1}(T-\muh_T)\leq r)
&=E\big[\si_T^{-1}(T-\mu_T)\leq r+Q(Y)|Y)\big]\\
&=E[\Phi(r+Q(Y))]\\
&=\Phi(r)+\phi(r)E[Q(Y]-\frac12r\phi(r)E[Q(Y)^2]\\
&\ \ \ +\frac12E\left[\int_r^{r+Q}(r+Q-x)^2(x^2-1)\phi(x)dx\right]\\
&\equiv \Phi(r)+\phi(r)T_1-\frac12r\phi(r)T_2+T_3
\end{align*}
Because $|r+Q-x|\leq |Q|$ for $x\in (r,r+Q)$ and $(x^2-1)\phi(x)$ is
uniformly bounded,
\begin{equation*}
T_3(r) = \frac12E\left[\int_r^{r+Q}(r+Q-x)^2(x^2-1)\phi(x)dx\right]
\leq C E\left[Q^2\int_r^{r+Q}dx \right]\leq CE |Q|^3
\end{equation*}
for some constant $C>0$.
Thus, the evaluation of $P(\sih_T^{-1}(T-\muh_T)\leq r)$ reduces to the evaluation of $EQ$, $EQ^2$ and $E|Q|^3$.
In particular, if we obtain $EQ=O(m^{-1})$, $EQ^2=O(m^{-1})$ and $EQ^8=O(m^{-4})$, then it follows that $E|Q|^3 = O(m^{-3/2})$ by Jensen's inequality so that 
\begin{align}
\label{eqn:1}
P(\sih_T^{-1}(T-\muh_T)\leq r)
=\Phi(r)+ O(m^{-1})\gamma(r,\beta,\psi)+O(m^{-3/2}),
\end{align}
where $\gamma$ is a smooth function of $O(1)$.
Because we consider the parametric bootstrap, the mathematical argument leading to the last display similarly yields 
\begin{equation*}
    P(\sih_T^{-1*}(T^*-\muh_T^*)\leq r)
    = \Phi(r) +O(m^{-1})\gamma(r,\widehat{\beta},\widehat{\psi})+ O(m^{-3/2})
\end{equation*}
with some appropriate modifications.
In the following, we provide a sketch of the proof of (\ref{eqn:1}) by verifying $EQ^8=O(m^{-4})$.
Once we obtain this result, proving the statement on the confidence interval is straightforward as in \cite{CLL2008}.

To analyze the moment of $Q$, first consider the element of $\muh_T-\mu$. We have
\begin{eqnarray*}
\widehat{\mu}_i - \mu_i&=&
A(A+D_i)^{-1}y_i+D_i(A+D_i)^{-1}X_i\beta 
-\widehat{A}(\widehat{A}+D_i)^{-1}y_i+D_i(\widehat{A}+D_i)^{-1}X_i\widehat{\beta} \\
&=&D_i(A+D_i)^{-1}X_i(X^t\Sigma^{-1}X)^{-1}X^t\Sigma^{-1}(v+\epsilon)\\
&&+D_i(A+D_i)^{-1}X_i\{(X^t\widehat{\Sigma}^{-1}X)^{-1}X^t\widehat{\Sigma}^{-1}-(X^t\Sigma^{-1}X)^{-1}X^t\Sigma^{-1}\}(v+\epsilon)\\
&& + \left(\widehat{A}(\widehat{A}+D_i)^{-1}-A(A+D_i)^{-1}\right)(J_i-X_i(X^t\widehat{\Sigma}^{-1}X)^{-1}X^t\widehat{\Sigma}^{-1})(v+\epsilon)\\
&&+\left(D_i(\widehat{A}+D_i)^{-1}-D_i(A+D_i)^{-1}\right)X_i\beta\\ &&
+\left(\widehat{A}(\widehat{A}+D_i)^{-1}-A(A+D_i)^{-1}\right)X_i\widehat{\beta}\\
&&=R_{1i}+R_{2i}+R_{3i}+R_{4i}+R_{5i}
\end{eqnarray*}
where $J_i$ is a diagonal matrix with 1 in the $j$th element with $j=m(i-1)+1,\ldots,mi$ and 0 otherwise. 
Let $R_i=(R_{i1}^t,\ldots,R_{im}^t)^t,i=1,\ldots,5$.
Thus, we can write 
\begin{equation*}
    Q(Y)  = \sigma^{-1}_T\left\{c^tR_1+c^tR_2+c^tR_3+c^tR_4+c^tR_5+q(\widehat{\sigma}_T-\sigma_T)\right\}.
\end{equation*}

We evaluate moments of $c^tM_1$. Clearly, $E[c^tR_1]=0$. For the second moment, a general term of the matrix $E[R_1R_1^t]$ is 
\begin{eqnarray*}
E[R_{1i}R^t_{1j}]
=D_i(A+D_i)^{-1}X_i(X^t\Sigma^{-1}X)^{-1}X_j^t(A+D_j)^{-1}D_j^t.
\end{eqnarray*}
Because $(X^t\Sigma^{-1}X)^{-1}=O(m^{-1})$ and $c$ is fixed, we obtain $E(c^tR_1)^2 = O(m^{-1})$. 
For the 8th moment, note that the 8th moment of the sum is the sum of the 8th moments up to constant.
Thus we consider the 8th moment of $c_jR_1{j}$ where $c=(c_1^t,\ldots,c_m^t)^t$ with $c_i\in\mathbb{R}^d,i=1,\ldots,m$.
Let $P_X = X(X^t\Sigma^{-1}X)^{-1}X^t\Sigma^{-1}$ and $I_i\in\mathbb{R}^{q\times mp}$ be a block matrix with blocks of zero matrices and one identity matrix such that $D_iX = I_i$. Because $P_XP_X=P_X$, it follows from the Cauchy-Schwartz inequality that
\begin{eqnarray*}
&&E(c_i^tR_{1i})^8
=E\left\{c_i^tD_i(A+D_i)^{-1}I_iP_XP_X(v+\epsilon)\right\}^8\\
&&\leq E\left\{c_i^tD_i(A+D_i)^{-1}I_iP_XP_X^tI_i^t(A+D_i)^{-1}D_ic_i (v+\epsilon)^tP_X^tP_X(v+\epsilon) \right\}^4.
\end{eqnarray*}
Because $c_i$ is fixed and $P_XP_X^t = O(m^{-2})$, 
\begin{eqnarray*}
E(c_i^tR_{1i})^8
\leq CO(m^{-8})E\{(v+\epsilon)^tP_X^tP_X(v+\epsilon)\}^4
\end{eqnarray*}
for some constant $C>0$.
Since $P_X=O(m^{-1})$ and $\{P_X(v+\epsilon)\}^2$ is the sum of $m$ terms, the above expectation is $O(m^4/m^2)=O(m^2)$.
Hence we obtain $E(c_i^tR_{1i})^8 = O(m^{-6})$.

To evaluate the 8th moment of the rest of terms in $Q(Y)$, we need to evaluate the moment of $(\widehat{A}+D_i)^{-1}-(A+D_i)^{-1}$ and $\widehat{\Sigma}^{-1}-\Sigma^{-1}.$
To see this, we have, for example, that
\begin{eqnarray*}
&&R_{2i}\\
&&=D_i\{(\widehat{A}+D_i)^{-1}-(A+D_i)^{-1}\}\\
&&\quad \times X_i\{(X^t\widehat{\Sigma}^{-1}X)^{-1}-(X^t\Sigma^{-1}X)^{-1}\}X^t(\widehat{\Sigma}^{-1}-\Sigma^{-1})(v+\epsilon)\\ 
&& \quad +D_i\{(\widehat{A}+D_i)^{-1}-(A+D_i)^{-1}\}
\\&&\qquad \times 
X_i\{(X^t\widehat{\Sigma}^{-1}X)^{-1}-(X^t\Sigma^{-1}X)^{-1}\}X^t\Sigma^{-1}(v+\epsilon)\\ 
&&\quad + D_i\{(\widehat{A}+D_i)^{-1}-(A+D_i)^{-1}\}
X_i(X^t\Sigma^{-1}X)^{-1}X^t(\widehat{\Sigma}^{-1}-\Sigma^{-1})(v+\epsilon)
\\&&\quad + D_i\{(\widehat{A}+D_i)^{-1}-(A+D_i)^{-1}\} X_i(X^T\Sigma^{-1}X)^{-1}X^t\Sigma^{-1}(v+\epsilon)\\
&&\quad +D_i(A+D_i)^{-1} X_i\{(X^t\widehat{\Sigma}^{-1}X)^{-1}-(X^t\Sigma^{-1}X)^{-1}\}X^t(\widehat{\Sigma}^{-1}-\Sigma^{-1})(v+\epsilon)\\ 
&&\quad +D_i(A+D_i)^{-1} X_i\{(X^t\widehat{\Sigma}^{-1}X)^{-1}-(X^t\Sigma^{-1}X)^{-1}\}X^t\Sigma^{-1}(v+\epsilon)\\ 
&&\quad +D_i(A+D_i)^{-1}
 X_i(X^t\Sigma^{-1}X)^{-1}X^t(\widehat{\Sigma}^{-1}-\Sigma^{-1})(v+\epsilon).
\end{eqnarray*}
Note that the evaluation of  $(\widehat{A}+D_i)^{-1}-(A+D_i)^{-1}$ and $\widehat{\Sigma}^{-1}-\Sigma^{-1}$ involves asymptotic expansions of matrix entries.
As pointed out by \cite{CLL2008} on page 1240, this computation involves several hundreds pages of elementary calculations.
In the end, both 8th moments reduce to the 8th moment of the Frobenius norm of  $\widehat{A}-A$ which is $O(m^{-4})$. We omit these details and refer to \cite{CLL2008}.

\vspace{1cm}

\bibliographystyle{chicago}
\bibliography{refs}

\end{document}